\documentclass[twocolumn,floatfix]{revtex4}
\usepackage{epsfig}

\begin{document}
\title{A lattice of Magneto-Optical and Magnetic traps for cold atoms}
\author{Axel Grabowski and Tilman Pfau}
\affiliation{5.Physikalisches Institut, Universit{\"a}t Stuttgart,
Pfaffenwaldring 57, 70550 Stuttgart, Germany}
\date{\today}
\begin{abstract}
We describe basic periodic trapping configurations for ultracold
atoms above surfaces. The approach is based on a simple wire grid
and can be scaled to provide large arrays of periodically arranged
magnetic or magneto-optical traps. The unit cells of the trap
lattices are based on crossed wire segments. By alternating the
current directions in the wires of the grid it can be
distinguished between 3 basic lattice configurations. As a first
demonstration, we used macroscopic wires in a 2 layer
configuration to realize the unit cells of the lattices. With this
experimental setup, we observe two of the basic unit cells and an
array of 2x2 magneto optical traps.
\end{abstract}
\maketitle

\section{Introduction}
\label{intro} An array of individual microtraps for ultracold
atoms, where each trap can be addressed separately, provides a
large number of new degrees of freedom for coherent manipulation
and control of cold atoms. Besides applications in Quantum
Computing, where  every lattice site corresponds to a qubit, an
addressable array will allow the dynamic definition of quantum
dots and quantum wires to perform e.g. transport measurements on
fully controllable structures filled with quantum gases.\\
One possible way to reach this goal could be microstructured
traps. Since the first proposals for such traps for neutral atoms
\cite{Weinstein} a lot of progress has been made in this field
\cite{Reichel1,H"ansel,Cassettari}, including the generation of
Bose Einstein Condensates (BEC) in microtraps \cite{Ott,H"ansel1}.
The main advantages of miniaturized traps are on the one hand the
possibility to create nearly arbitrary potentials and on the other
hand the possibility to reach already with small currents traps
with much tighter confinement than in large macroscopic traps.
Reviews on the basic principles of such microtraps can be found in
references \cite{Schmiedmayer2,Reichel}.

In this paper, we present basic periodic trapping configurations
which can be realized with multilayer microstructures. The
trapping fields for the atoms (Ioffe Pritchard type potentials for
magnetic trapping of atoms as well as quadrupole fields for
magneto optical trapping) are formed by two layers of crossed
wires, which can be individually addressed. We present first
experiments with a system of macroscopic wires and show, that it
is possible to produce multiple magneto optical traps (MOT) next
to a surface in a controlled manner. This could be a first step to
load ultracold atoms into an array of microstructured magnetic
traps in a parallel way. We envisage that experiments with trapped
atoms or even the production of BECs in such arrays of traps can
be performed.

The paper is organized as follows: In section
\ref{Basicconfigurations} a review of the basic concepts of
trapping atoms with a single wire, respectively with
microstructures is given. In part \ref{Latticeconfigurations} we
present a scalable implementation of trapping configuration for
cold atoms in periodic arrangements. In section \ref{Experimental
setup} the experimental setup is introduced and finally, in
section \ref{Experimental results} we present first experiments
including multiple MOT-systems based on such multilayer
structures.

\section{Configurations}
\label{sec:1}
\subsection{Basic configurations} \label{Basicconfigurations}
\begin{figure}[h!]
\begin{center}
\epsfig{file=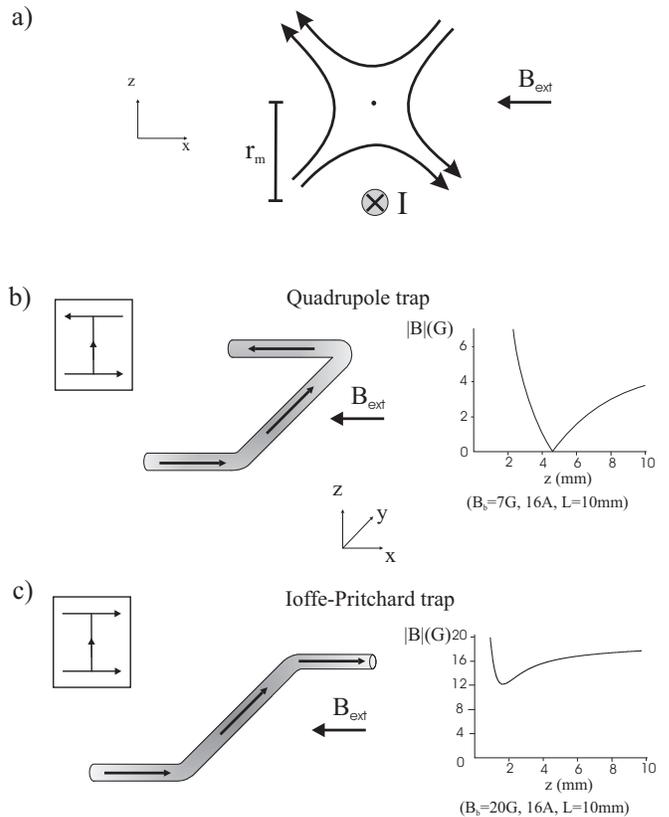,width=\columnwidth} \caption{Different
wiretrap configurations: a) The superposition of the magnetic
field produced by a wire with an external homogeneous bias field
forms a two dimensional quadrupole with open ends. b) Bending of
the wire in a ''U'' form gives a quadrupole trap. c) Bending of
the wire in a ''Z'' form gives a Ioffe-Pritchard type trap. The
graphs show for both configurations the magnetic field in the
middle of the the two bent parts above the wire. The length L of
the central bar is in both cases 10 mm, the current I in the wire
16A and the external bias field 7 G respectively 20 G. The insets
in b,c) show the corresponding ''H'' configurations.} \label{fig1}
\end{center}
\end{figure}

The simplest configuration to produce magnetic potentials which
can be used to catch cold atoms is the superposition of the
magnetic field of an infinitely long wire
$B=|\vec{B}|=\frac{\mu_{0}}{2 \pi}\frac{I}{r}$ ($I$ is the current
in the wire, $r$ the distance from the wire) with an external
homogenous bias field $B_{b}=|\vec{B}_b|$ perpendicular to the
wire \cite{Frisch} (see fig. \ref{fig1} a). This configuration
yields a two dimensional quadrupole field parallel to the wire
with a field minimum at a distance of $r_{m} =\frac{\mu_{0}}{2
\pi}\frac{I}{B_{b}}$. Next to this minimum the magnetic field
increases linearly in all directions perpendicular to the wire
with a gradient of $\left.\frac{dB}{dr}\right|_{r_{m}} =-\frac{2
\pi}{\mu_{0}}\frac{B_{b}^{2}}{I}$ \cite{Reichel1}. This type of
trap can be extended to a harmonic trap by applying an external
field $B_{p}=|\vec{B}_p|$ parallel to the wire. The resulting
curvature in the radial direction at the height $r_{m}$ can be
expressed as
\begin{equation}
\left.\frac{d^2B}{dr^2}\right|_{r_{m}} = \left(\frac{2
\pi}{\mu_{0}}\right)^2 \frac{B_{b}^4}{B_{p} I^2}
\end{equation}
where I is the current in the wire \cite{Schmiedmayer1}. For this
one wire configuration, the bias field $B_ {b}$ can alternatively
be produced by two additional wires on both sides of the trapping
wire with currents in opposite direction \cite{Thywissen}. A
similar trapping configuration consisting of four wires in
parallel is possible too \cite{Thywissen}.

The 2-dimensional trapping potential of infinitely long wires can
be extended to real 3-D trapping configurations by closing the
quadrupoles in the third dimension by adding two wires
perpendicular to the main wire. This can be done with one wire by
bending it in so called ''U ''or ''Z'' configurations (for
illustration see fig. \ref{fig1}b,c) \cite{Reichel1}, resulting in
3-D quadrupoles (U-configuration) or Ioffe-Pritchard type traps
(Z-configuration) respectively.

A 3-D magnetic quadrupole has three main axes which go through the
magnetic zero and are orthogonal with respect to each other. To
satisfy $\vec{\nabla} \cdot \vec{B}=0$, there is one axis where
the sign of the field gradient must be different than for the
other two. For the further explanation we use this axis to define
the orientation of the 3-D quadrupole in space. Furthermore we
define the bent parts of the wire to be parallel to the
x-direction, where the central part is in the y-direction. The
z-direction is perpendicular to the plane defined by the x-and
y-axes.

In the ''U'' configuration \cite{Reichel1} the y-components of the
magnetic field of the bent wires cancel out at the origin of the
linear quadrupole (see fig. \ref{fig1}b), whereas the z-components
add. Due to this the origin of the quadrupole is shifted in the +x
direction away from central bar. The gradients in both the
x-direction and the y-direction are
\begin{eqnarray}
\left. \frac{dB}{dx} \right|_{r_{m}} &=&-\frac{\mu_{0}}{2 \pi} \frac{I_{w}}{r_{m}^2} \label{gra1}\\
\left. \frac{dB}{dy} \right|_{r_{m}}&=&\frac{2 \mu_{0}}{\pi}
\frac{I_{c} r_{m}L}{(L^2 + 4 r_{m}^2)^2}, \label{gra2}
\end{eqnarray}
where $r_m$ is the distance of the quadrupole minimum from the
wire, L is the length of the central bar of the U-trap and
$I_{w}=I_{c}=I$ is the current in the wire. In this formula the
shift due to the bent parts is not taken into account. This
simplification is valid, if $r_m$ is small compared to the spacing
$L$ between the bent parts of the wire. The axis of the quadrupole
in the configuration shown in fig. \ref{fig1}b is lying in the
x-z-plane above the central bar between the bent parts of the wire
under an angle of $45^{\circ}$ with respect to the x-axis. The
bent parts of the ''U'' configuration do not close the quadrupole
in a symmetric way. A more symmetric configuration is a ''H''
configuration, which can not be bent in a simple way with a single
wire (see inset fig. \ref{fig1}b).

In the ''Z''-configuration \cite{Reichel1} a Ioffe-Pritchard-type
trap is formed (see fig. 1c). In this configuration the
y-components of the magnetic field of the bent parts add and the
z-components cancel out in the middle of the central bar. For this
reason a magnetic field minimum appears in the middle of the
central bar at the distance $r_m$. Due to the non vanishing
y-components the magnetic field at the field minimum is non zero
and can be calculated as
\begin{equation}
B_{m}=\frac{2 \mu_{0}}{\pi} \frac{I_{c} r_{m}}{L^2 + 4 r_{m}^2},\label{xyz}\\
\end{equation}
where L is the length of the central bar and $I$ is the current in
the wire. The curvature of these traps at the minimum $r_{m}$ are
in both the x-direction and the y-direction
\begin{eqnarray}
\left.\frac{d^2B}{dx^2}\right|_{r_{m}} &=&\frac{\mu_{0}}{16 \pi}
\frac{I_{w} (L^2 + 4 r_{m}^2)}{r_{m}^5} \label{curv1} \\
\left.\frac{d^2B}{dy^2}\right|_{r_{m}}&=&-\frac{2 \mu_{0}}{\pi}
\frac{I_{c} (16r_{m}^4-16r_{m}^2L^2-L^4)}{r_{m} (4r_{m}^2+L^2)^3}
\label{curv2}
\end{eqnarray}
This kind of trap is more stable against Majorana transitions than
the quadrupole trap and is thus favourable for magnetic trapping
of atoms, whereas ''U'' type traps are favourable for
magneto-optical trapping of atoms, which typically requires a
quadrupole configuration. This ''Z'' trap is again asymmetric and
can be extended to an ''H'' form trap, which is a more symmetric
situation (see inset fig. \ref{fig1}c).

\subsection{Lattice configurations} \label{Latticeconfigurations}

\begin{figure}[h!]
\begin{center}
\epsfig{file=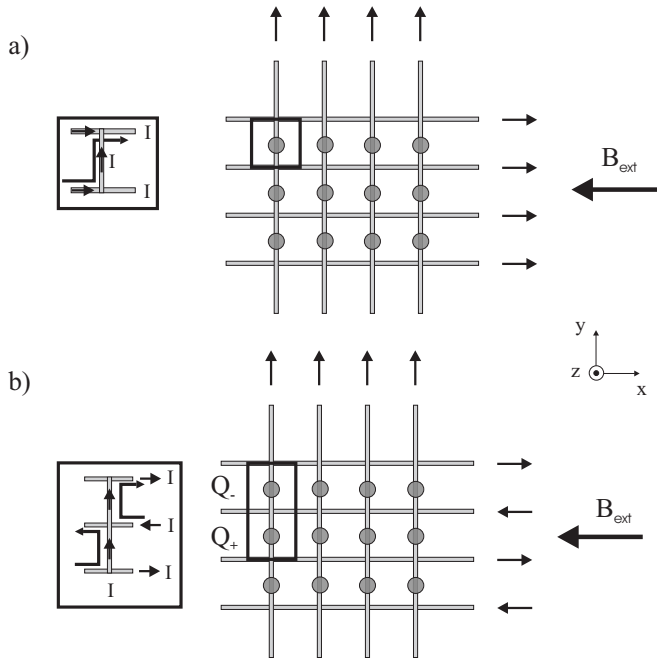,width=\columnwidth} \caption{ Basic
configuration for the trapping of atoms in a periodic structure
with 2 layers of wires. a)  The currents in the layers are flowing
in all wires in same direction. An additional bias field produces
an array of Ioffe Pritchard type traps. The inset on the left side
shows the unit cell in this lattice configuration (Configuration
A). b) The currents in one layer are all flowing in the same
direction, in the second layer the currents are bidirectional.
With an additional bias field one gets an array of quadrupoles
($Q_{-}$,$Q_{+}$), which have different orientations of their
quadrupole axis. The inset on the left side shows the unit cell in
this lattice configuration (Configuration B).} \label{fig2}
\end{center}
\end{figure}

\begin{figure}[h!]
\begin{center}
\epsfig{file=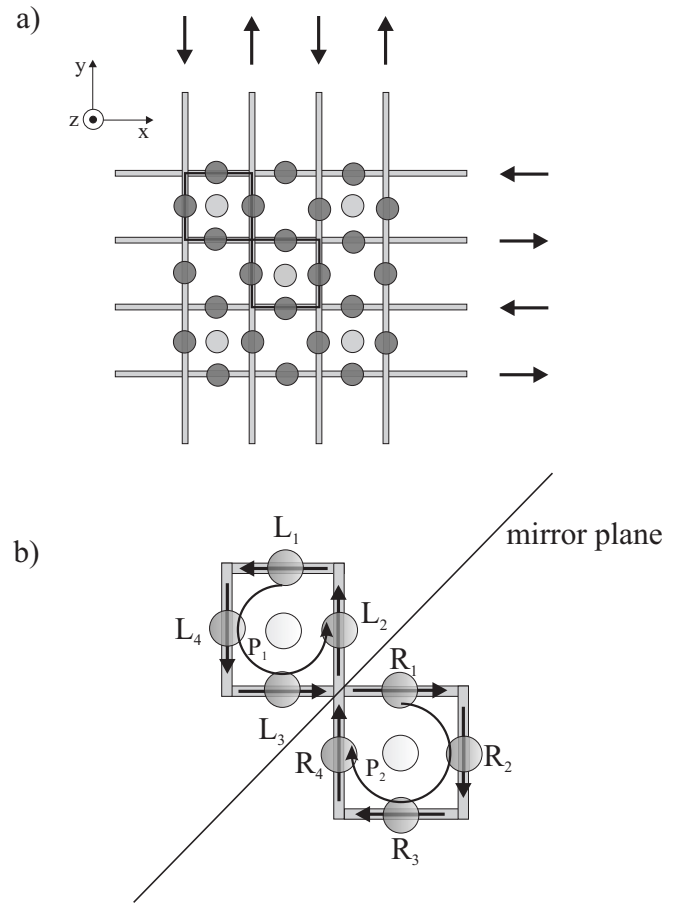,width=\columnwidth} \caption{a) Lattice
configuration with bidirectional currents in both layers. This
gives a rich structure of quadrupoles in different directions. b)
The simple unit cell for this array of wires consists of 6 wires.
Without an external bias field this configuration yields 10
different quadrupoles as explained in the text (Configuration C).}
\label{fig3}
\end{center}
\end{figure}

To extend these basic schemes to an array of traps, let us
consider 2 layers of parallel wires, where the 2 layers are
perpendicular to each other (see fig. \ref{fig2} and fig.
\ref{fig3}). For simplification let us assume that all wires carry
the same constant current I, the wires are infinitely thin and
infinitely long. In addition to the magnetic field produced by
these wires an external bias field $B_{b}$ can be applied parallel
or perpendicular to the 2 layers. Such a configuration is very
versatile and allows to create many different lattice geometries.
The possible lattice configurations are built up from simple unit
cells, which will now be described in detail.

We consider three different configurations. In the first
configuration, the currents $I$ in each of the layers are
unidirectional (see fig. \ref{fig2}a, configuration A). In the
second configuration the currents are unidirectional in one layer
(y-direction), and alternating in the second layer (x-direction)
(see fig. \ref{fig2}b, configuration B). In the last configuration
the currents are bidirectional in both layers (see fig.
\ref{fig3}a, configuration C).

Configuration A (currents in both layers unidirectional) has
already been discussed in references \cite{Reichel2,Yin}. We
describe the concept using simple unit cells consisting of 3 wires
segments crossing each other as marked in the box in fig.
\ref{fig2}a. This unit cell has a central bar in the middle,
enclosed by two wire segments in the upper and lower part of the
unit cell. The size of the unit cell is given by the distance
between the wires in each layer. The lattice is a simple square
lattice. By applying an external bias field in the x-direction as
indicated in fig. \ref{fig2}a each ''H'' shaped unit cell contains
a Ioffe Pritchard trap above the central bar. The basic properties
of an individual trap can be calculated from the equations
\ref{curv1} and \ref{curv2} given above for the ''Z''-trap plus a
perturbation term, which will be calculated later.

In configuration B, the currents are unidirectional in one layer
(y) and alternating in the second layer (x) (see fig \ref{fig2}
b). Here the simple unit cell consists of 4 wire segments as shown
in fig. \ref{fig2} b). The lattice is rectangular. By applying an
external bias field in the x-direction two different quadrupole
traps $Q_{-}$,$Q_{+}$, (see fig. \ref{fig2}b) are contained above
the surface in the unit cell, which can be described by the
equations \ref{gra1} and \ref{gra2} given for a quadrupole trap
plus a perturbation term. The difference between the two
quadrupoles in the unit cell is the orientation. At quadrupole
$Q_{-}$ the axis is lying in the x,z plane under an angle of
$-45^{\circ}$ with respect to the x-axis, and at quadrupole
$Q_{+}$ the axis is under an angle of $45^{\circ}$ with respect to
the x-axis. The quadrupoles can be used for magneto optical
trapping to catch atoms in a reflection MOT
\cite{Reichel2,Lee,Schneble}. Due to the fact that the two
quadrupoles have different orientations, the circular polarisation
of the MOT beams has to be different to either use the upper or
the lower of the quadrupoles as will be shown in the experimental
section.

In configuration C, the currents are alternating in both layers
(fig. \ref{fig3}a). The unit cell consists of 8 wire segments.
Even without an external bias field in this cell 10 different
quadrupoles can be found. The bias field for all the different
quadrupoles is given by the neighboring wires as mentioned before
\cite{Thywissen}, so no external bias field is necessary. Eight of
the quadrupoles can be found above the wires ($R_{1}$..$R_{4}$ and
$L_{1}$..${L_4}$, see fig. \ref{fig3}). The quadrupole axes of all
of them are tilted by $45^{o}$ against the surface with respect to
the wires and can be used for a reflection MOT
\cite{Reichel2,Lee,Schneble}. Two other quadrupoles ($P_1$,$P_2$)
are built up by two concentric wire squares \cite{Thywissen}. The
quadrupole axes of these two are oriented perpendicular to the
surface, and thus they can not be used for a reflection MOT. The
unit cell itself is symmetric under a mirror operation around the
mirror plane as indicated in fig. \ref{fig3} b). The lattice is a
simple square lattice.

In all three configurations, the magnetic field $\vec{B}$ in the
unit cell is given by the magnetic field produced by the wires in
the unit cell $\vec{B}_{c}$ plus a perturbation term from the rest
of the lattice $\vec{B}_{p}$
\begin{equation}
\vec{B}=\vec{B}_{c}+\vec{B}_{p}
\end{equation}
For configuration A the perturbation in one cell can be expressed
by the formula
\begin{eqnarray}
\vec{B}_{p}& =& \frac{\mu_{0}I}{2 \pi}\sum_{\begin{array}{c} n=-\infty \\
n\neq 0 \end{array}}^{\infty}\frac{1}{(x-nd)^2+z^2}\cdot \left(
\begin{array}{c}  x-nd \\ 0 \\ -z \end{array} \right) \nonumber \\
&+& \frac{\mu_{0}I}{2 \pi}\sum_{\begin{array}{c} m=-\infty \\
m\neq 0 \end{array}}^{\infty}\frac{1}{(y-md)^2+z^2}\cdot \left(
\begin{array}{c}  0 \\ y-md \\ -z  \end{array} \right)
\label{pert1}
\end{eqnarray}
and for configuration B
\begin{eqnarray}
\vec{B}_{p} &= &\frac{\mu_{0}I}{2 \pi}\sum_{\begin{array}{c} n =-\infty \\
n \neq  0 \end{array}}^{\infty}\frac{(-1)^n}{(x-nd)^2+z^2}\cdot
\left(\begin{array}{c}  x-nd \\ 0 \\ -z \end{array} \right) \nonumber \\
&+& \frac{\mu_{0}I}{2 \pi}\sum_{\begin{array}{c} m=-\infty \\
m\neq 0 \end{array}}^{\infty}\frac{1}{(y-md)^2+z^2}\cdot \left(
\begin{array}{c}  0 \\ y-md \\ -z \end{array}\right)
\label{pert2}
\end{eqnarray}
where d is the distance between the wires and I the current
flowing in the wires. The sums over n and m runs over all wires in
the two horizontal directions. The terms for $n=m=0$ are excluded
from the sum since only the perturbation from the rest of the
lattice on a given unit cell should be considered here. An
analogous situation is given in the third configuration. The
perturbations to the magnetic field in one lattice cell will be
small, if $d$ is much larger than the distance $z=r_{m}$ of the
potential minimum from the surface.

For finite systems and $r_{m} \gg\ d$ the sum above is finite and
one has to consider edge effects, which perturb the symmetry. This
is analogous to finite-size magnetic mirrors used in atom optics
\cite{Lau}. The effect can be compensated by adding compensation
wires, as proposed in \cite{Zabov}.

\section{Experimental setup}\label{Experimental setup}

\begin{figure}[h!]
\begin{center}
\epsfig{file=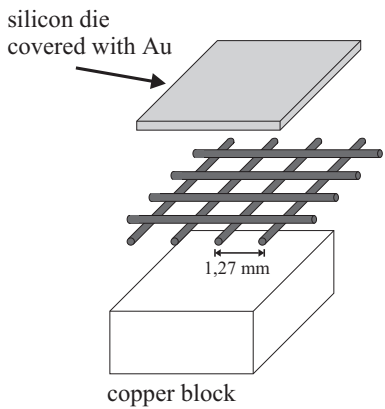,width=5cm}
 \caption{Experimental setup inside the vacuum chamber. Above a
copper block 2 layers of kapton insulated wires are mounted under
a gold covered silicon plate. The whole setup is mounted upside
down inside the vacuum chamber.} \label{fig4}
\end{center}
\end{figure}

To test the configurations proposed in section
\ref{Latticeconfigurations} arrays of magnetic quadrupole traps
are created by 2 layers of macroscopic wires mounted inside a
vacuum chamber (see fig. \ref{fig4}). Each layer of wires consists
of a ribbon UHV cable with a pitch of 1.27 mm (Caburn, Kapton
Ribble Wire). The individual wires are kapton insulated with a
diameter of the copper core of 0.25 mm. The cables are glued
together with UHV compatible epoxy and are mounted on top of each
other in a two layers configuration, where the upper layer
consists of 10 wires in parallel and the lower layer consists of 8
parallel wires below the first one. The distance between the
center of the wires in the two layers is 1.27 mm, given by the
wire diameter. The whole wire assembly is attached to a Cu holder.
The maximum current in the wires was about 5A. At higher currents,
the pressure in the vacuum chamber started to increase probably
due to outgasing of the insulating kapton layer. On top of the
upper layer of wires a Si plate (thickness: 0.4 mm) covered with
10nm of Cr and 100 nm of Au is mounted, which serves as mirror.
The whole setup is mounted upside down inside the chamber. The
mirror at the surface is necessary, because we use the reflection
MOT principle \cite{Reichel2,Lee,Schneble} to catch cold $^{87}Rb$
atoms. For this type of MOT two of the laser beams are parallel to
the surface and two other lasers beams are reflected from the
surface under an angle of $45^{\circ}$. The magnetic quadrupole
field necessary for the MOT is generated by the two layers of
wires plus an external bias field. This bias field can be adjusted
between 0 G and 80 G. The circular polarisation of the four beams
depends on the orientation of the magnetic quadrupole. The
$^{87}Rb$ is provided by a Rb dispenser \cite{SAES,Fortagh}
mounted approximately 4 cm away from the assembly. The optical
access is provided by a large window below the sample.

To operate the MOT, we use as lasers two home built grating
feedback stabilized diode lasers, operating at 780 nm. The cooling
laser is locked on the $5S_{1/2},F=2 \rightarrow 5P_{3/2},F'=3$
cycling transition using polarisation spectroscopy in a vapor
cell. The laser power in this laser is about 17 mW. The repumping
laser is locked to the $5S_{1/2},F=1 \rightarrow 5P_{3/2},F'=2$
transition using the same locking scheme as for the MOT laser.
This laser has a total power of 7 mW.

To detect the atoms in the MOT, we used a calibrated 8-bit CCD
Camera, placed 8 cm under the surface outside the vacuum chamber.
The imaging system is set up in a way, that atoms above the whole
surface inside the chamber can be detected. One pixel on the
camera corresponds to a $39 \mu$m$\times$$36\mu m$ area in the
trap region.

\section{Experimental results}\label{Experimental results}

\begin{figure}[h!]
\begin{center}
\epsfig{file=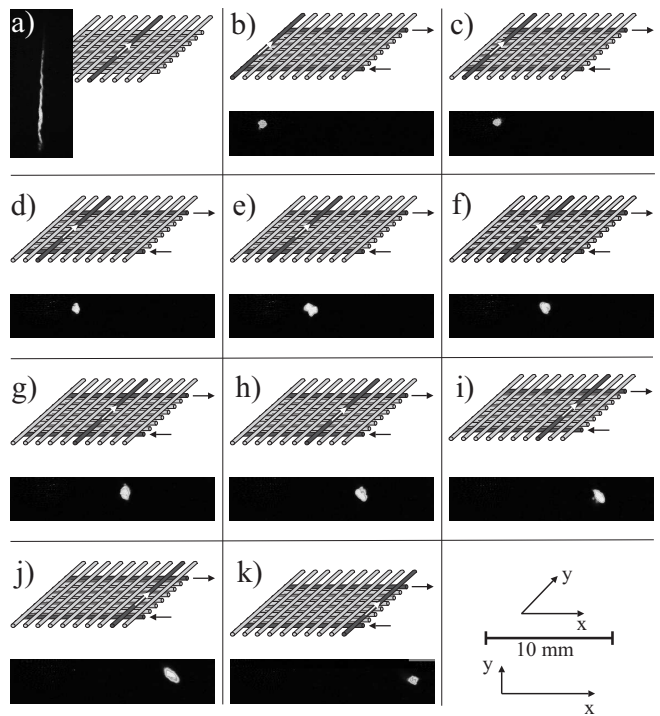,width=\columnwidth}  \caption{a)
Two-dimensional MOT produced by the magnetic field of a single
wire with a current of 4 A and an external bias field of 3 G. All
MOT beams are turned on yielding a 2-D MOT with an optical
molasses on the y-axis. b-k): MOTs with a magnetic quadrupole
field produced by two layers of wires. By turning on the current
in two wires in the x-direction MOTs at different position above
the surface can be addressed by using different wires. The wires
used are indicated above the images. The currents in the wires
were 4 A with an external bias field of 3 G. The distance between
the wires in the x-direction is 1.27 mm (left to right in the
picture), the distance between the current-carrying wires in the
y-direction is 10.2 mm.} \label{fig5}
\end{center}
\end{figure}

By using one wire in the upper one of the two layers together with
the external bias field and turning on all the MOT beams one has
the situation of a 2D MOT with additional optical molasses on the
axis parallel to the surface (see fig. \ref{fig5}), upper left
corner). This 2D-MOT can be extended to a 3D-MOT by using the
wires of the lower layer transform the 2-D quadrupole to an 3-D
quadrupole. In the experiments we used in the lower layer the two
outermost wires of our experimental setup, which are at a distance
of 10.1 mm (see fig. \ref{fig5}).

We found that the position of the MOT can be precisely controlled
by choosing the current carrying wire in the upper layer. The
currents were set to 4 A and an external bias field of 3 G was
applied. With these values the magnetic field gradients be
calculated from equations \ref{gra1} and \ref{gra2} as
$\frac{dB}{dx}=11 G/cm$ and $\frac{dB}{dy}=1.9 G/cm$. The
calculated distance $r_m$ from the MOT to the surface is 1.6 mm.

In this configuration, a MOT was realized above every wire (see
fig. \ref{fig5}). The average number of atoms for the different
MOT positions was $3 \cdot 10^{5}$. A situation where no external
bias field was necessary was reached by using as bias field wires
the neighboring wires on the left and on the right of the MOT wire
in the upper layer with current flowing in the opposite direction.
The number of atoms in such a MOT generated without external field
was lower, due to the fact that the capture range was smaller. The
reason for this is the fact that the bias field produced by the
two neighboring wires is inhomogeneous. The quadrupoles resulting
from this configuration extend over a smaller volume and so the
capture range is reduced.

\begin{figure}[h!]
\begin{center}
\epsfig{file=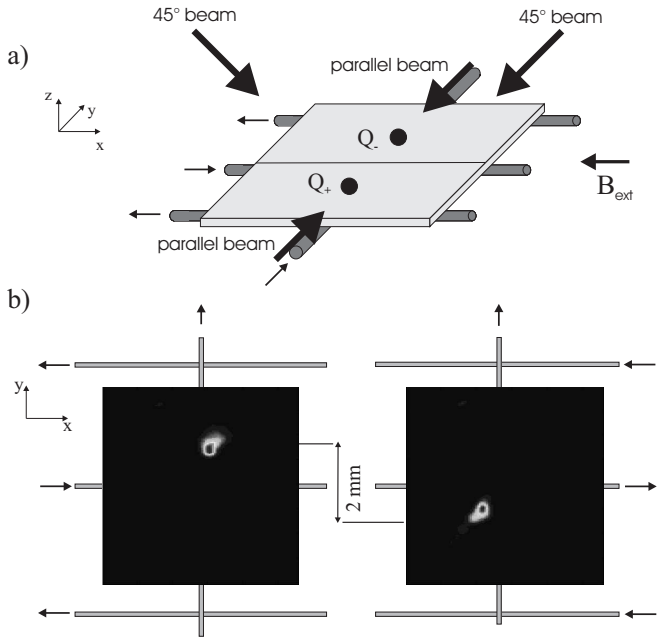,width=\columnwidth} \caption{a) Experimental
realization of the unit cell of configuration B. The wire
configuration is consists of 4 wire segments. In the
reflection-MOT configuration 2 MOT beams are reflected from the
surface and 2 beams are parallel to the surface (arrows). b)
Switching between the different MOTs by changing the circular
polarisation of the beam parallel to the surface. The current in
the wires was 4 A, the external bias field 3 G. The position
displacement by changing the polarisation is about 2 mm in the
y-direction and 0.2 mm in the x-direction. The change in position
in x-direction is probably due to an imbalance in the MOT beams.}
\label{fig6}
\end{center}
\end{figure}

\begin{figure}[h!]
\begin{center}
\epsfig{file=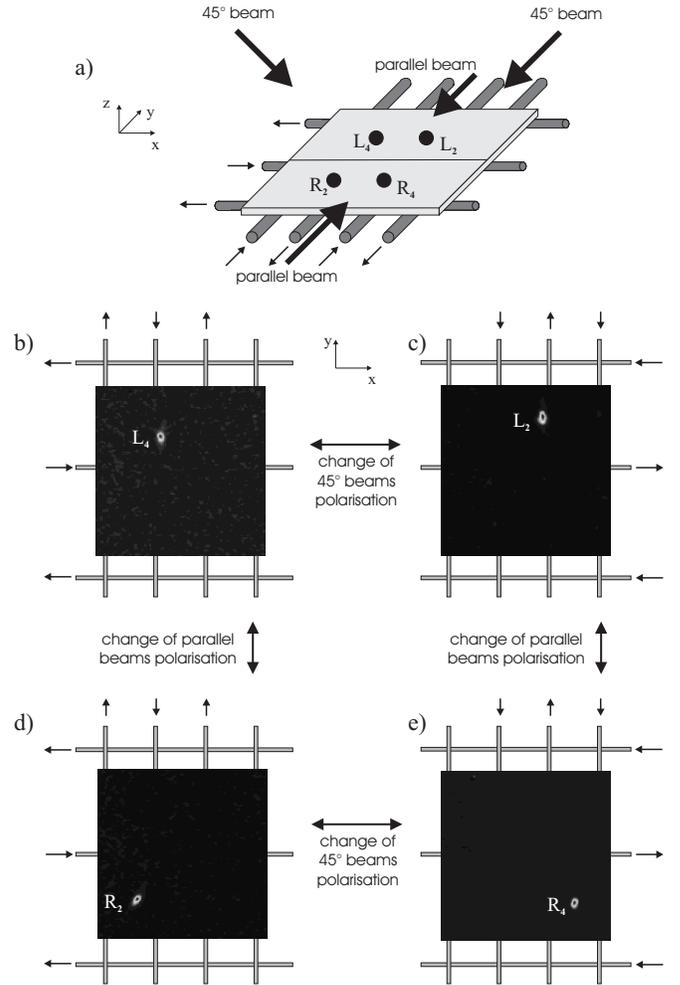,width=\columnwidth} \caption{a) Experimental
realization of the unit cell of configuration C. The wire
configuration consists of 7 wires, of which for the experiments
only 6 are used simultaneously. In the MOT configuration 2 MOT
beams are reflected from the surface and 2 beams are parallel to
the surface (arrows). b,c)-d,e) switching between the different
MOTs by reversing the circular polarisation of the beams parallel
to the surface. b,d)-c,e) switching between the MOTs by reversing
the circular polarisation of the $45^{\circ}$ MOT beams. The
current in the wires was 4 A, the external bias field 3 G.}
\label{fig7}
\end{center}
\end{figure}

In the proposed configuration B two quadrupoles which are
different in orientation emerge above the middle wire segment of
the unit cell. These two quadrupoles can be used to operate the
MOT, if the respective polarisations of the MOT laser beams fit to
the orientation of the quadrupoles. In our experimental setup, we
realized the unit cell configuration consisting of 4 wires (see
fig. 6a). The distance between the parallel wires was 5mm (upper -
middle wire), respectively 3.8 mm (middle - lower wire). The
current in each of the wires was chosen as 4 A and an external
bias field of 3 G was applied. This yields a gradient of
$\frac{dB}{dx}=11 G/cm$ in the x- and $\frac{dB}{dy}=3.3 G/cm$ in
the y-direction. The distance $r_m$ above the surface is again
$1.6 mm$. By changing the polarisation of the MOT beams parallel
to the surface from $\sigma^+$ to $\sigma^-$ the MOT changes its
position due to the two different orientations of the two
quadrupoles within the rectangular unit cell (see fig. \ref{fig6}
b).

In configuration C, 10 quadrupoles are created per unit cell
without external fields. As explained before 8 of these
quadrupoles are in principle suited to operate a reflection MOT,
because the angle between the quadrupole axis and surface is
$45^{\circ}$. In our experimental configuration only four of them
can be seen due to the fact that only for four quadrupoles the
axes coincide with the direction of a MOT beam. The accessible
positions can again be addressed by changing the polarisation of
the MOT beams, where the polarisation configuration is determined
by the orientation of the magnetic quadrupoles for the different
wire configurations (see fig. \ref{fig7}a). For the generation of
the magnetic field in the unit cell we used 3 wires in the
y-direction with alternating currents (distance wire-wire 2.5mm)
and 3 wires in the x-direction (distance between the wires 5 mm,
respectively 3.8 mm, as in configuration B). By building up only
one such lattice cell, the quadrupoles at the edges of the lattice
cells are not accessible, because only in the middle of the cell
the bias field needed is produced by the neighboring wires. So we
investigated in this configuration the quadrupoles in the middle
of the unit cell ($R_{2}$,$L_{4}$). Then we switched off the
current in the left wire, and turned on the current in the right
wire to be able to observe the quadrupoles $R_{4}$ and $L_{2}$
(see configuration marked in fig. \ref{fig7}b-e). To reach a
stable situation the current in the two middle wires was in both
cases 3.5 A (y-direction) and in the outermost wires 5A. These
values yield a distance of the quadrupole from the surface of 0.2
mm and a gradient $\frac{dB}{dx}=20 G/cm$ in the x-direction. The
currents in the x direction were 4A. This yields a gradient of
$\frac{dB}{dy}=9.7 G/cm$ in the y-direction. As shown in fig.
\ref{fig7}b-e the different quadrupoles can be addressed by
changing the polarisations of the MOT beams as expected. By
changing the circular polarisation of the beams parallel to the
surface we can switch between $L_4$ and $R_2$ (see fig.
\ref{fig7}b,c) and $L_2$ and $R_4$ (see. fig. \ref{fig7}d,e)
respectively. The distance change in the y-direction is 4.4 mm
respectively 4 mm and in the x-direction 1 mm. By changing the
circular polarisation of the $45^{\circ}$ beams (see fig.
\ref{fig7}b,d, c,e) one can switch between $L_{4}$ and $L_{2}$,
respectively between $R_{2}$ and $R_{4}$. Here the observed change
in position in x direction is 2.2 mm. The number of atoms in the
different MOTs is between $8\cdot 10^4$ and $2.4 \cdot 10^5$
atoms. The main advantage of this configuration compared to
configuration B is the fact that no external bias field in
necessary.

\begin{figure}[h!]
\begin{center}
\epsfig{file=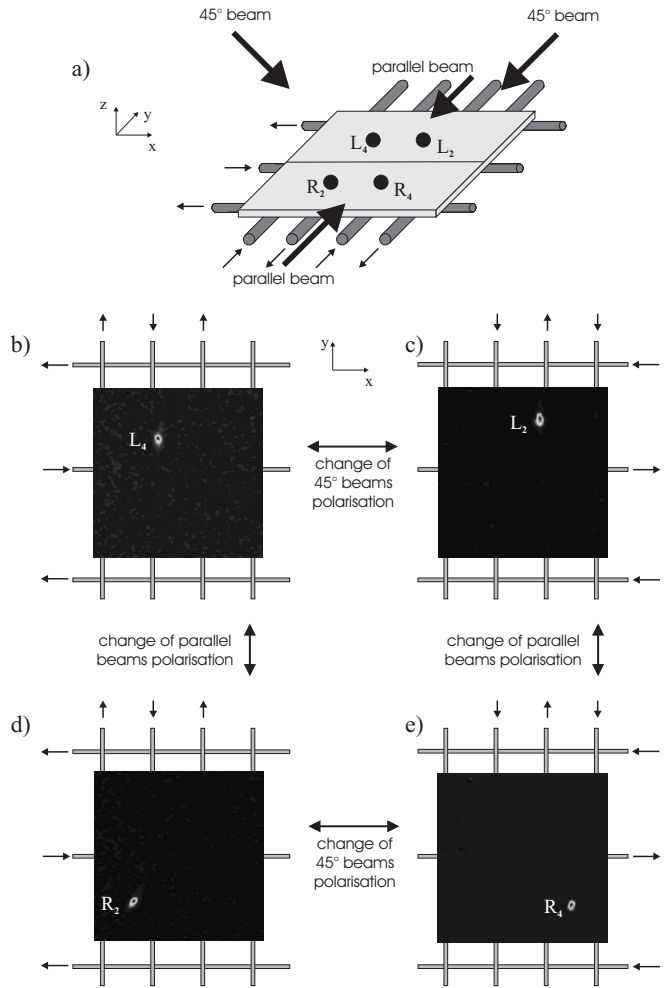,width=\columnwidth} \caption{a) Current
distribution in the lattice to observe an 2x2 array of MOTs. In
the y-direction 2 wires are used as indicated in the picture. In
the x-direction two neighboring wires are used as one effective
wire. b) 2x2 array of MOTs above the surface. By using 2 wires in
the y and 4 effective wires in the x-direction a stable situation
with 4 MOTs above the surface can be generated. The number of
atoms in each individual MOT is about $2 \cdot 10^5$ atoms. The
distance between the different MOTs is 3.3 mm in the y- and 8 mm
in the x-direction. c,d) By applying an additional laser beam to
the positions of the MOTs these can be individually turned on and
off by blowing the cold atoms away.} \label{Fig8}
\end{center}
\end{figure}

Up to now only single unit cells were observed. To built up
lattices it is necessary to create arrays of unit cells. This can
be done in the experiment by using more wires. For an array of
MOTs, we arranged 4 unit cells in configuration B. As explained
before it is necessary that the distance between the wires has to
be large compared to the distance of the field minimum $r_{m}$ to
the wire. To reach this, we used two wires in the x-direction with
a distance of 10.16 mm from each other. In the y-direction we used
two wires in pairs (current flowing in the same direction) as one
effective wire with a current of about 3 A in each of the
effective wires (see fig. \ref{Fig8}a). The complete configuration
in this direction is given by 4 of these wire pairs. The current
in the wires of the upper layer was also 3 A, the bias field was 4
G. With this configuration, we could generate a stable situation
with 4 MOTs operating above the surface (see fig. \ref{Fig8}b).
The gradient of the quadrupole in the x-direction is again
$\frac{dB}{dx}=20 G/cm$ and in the y-direction $\frac{dB}{dy}=4.3
G/cm$. The calculated distance $r_m$ above the surface is about
$1$ mm. The number of atoms in each individual MOT is about $2
\cdot 10^5$. The distance between the different MOTs is 3.3 mm in
the x- and 8 mm in the y-direction. To switch the MOTs
individually, we applied a laser pulse split off from the MOT
laser to turn the different MOTs on and off by blowing the atoms
away (see fig. \ref{Fig8}c,d). By turning off one MOT no
significant change in atom number of the remaining
MOTs could be detected.\\

\section{Summary}\label{Summary}

In this paper, we have proposed different lattice configurations
consisting of magnetic quadrupoles which are generated by 2
orthogonal layers of wires. These quadrupoles could be detected in
the different unit cells by operating MOTs with the required
polarisation configuration. We could furthermore demonstrate the
possibility to use this two layer configuration to trap atoms in a
2x2 MOT array above a surface. We were able to turn the MOTs on
and off individually by using an extra laser beam. Such a system
of multiple MOTs could be useful to load an array of magnetic
microtraps from an array of separate MOTs. The current number of
$10^{5}$ atoms in the single MOT will be increased in future
experiments by using the Rb dispenser in our chamber in a pulsed
operation with higher currents.

\section*{Acknowledgement}
We acknowledge support of the European Research Training Network
''Cold Quantum Gases'' under Contract No. HPRN-CT-2000-00125.\\
We thank Axel G\"orlitz for helpful discussions and proof reading.\\


\begin{thebibliography}{99}
\bibitem{Weinstein}
J.D. Weinstein and K.G. Libbrecht, \textit{Phys. Rev.}
\textbf{A52}, 4004 (1995).

\bibitem{Reichel1}
J. Reichel, W. H\"ansel and T.W. H\"ansch, \textit{Phys. Rev.
Lett.} \textbf{83}, 3398 (1999).

\bibitem{H"ansel}
W. H\"ansel, J. Reichel, P. Hommelhoff and T. W. H\"ansch,
\textit{Phys. Rev. Lett.} \textbf{86}, 608 (2001).

\bibitem{Cassettari}
D. Cassettari, B. Hessmo, R. Folman, T. Maier and J. Schmiedmayer,
\textit{Phys. Rev. Lett.} \textbf{85}, 5483 (2000).

\bibitem{Ott}
H. Ott, J. Fortagh, G. Schlotterbeck, A. Grossmann and C.
Zimmermann, \textit{Phys. Rev. Lett.} \textbf{87}, 230401 (2001).

\bibitem{H"ansel1}
W. H\"ansel, P. Hommelhoff, T. W. H\"ansch and J. Reichel,
\textit{Nature} \textbf{413}, 498 (2001).

\bibitem{Schmiedmayer2}
R. Folman, P. Kr\"uger, J. Denschlag, C. Henkel and J.
Schmiedmayer, \textit{Adv. At. Mol. Opt. Phys.} \textbf{48}, 263
(2002).

\bibitem{Reichel}
J. Reichel, \textit{Appl. Phys.} \textbf{B75}, 469 (2002).

\bibitem{Frisch}
R. Frisch and E.Segr\`e, \textit{Z. Phys.} \textbf{80}, 610
(1933).

\bibitem{Schmiedmayer1}
A. Haase, D. Cassettari, B. Hessmo and J. Schmiedmayer,
\textit{Phys. Rev.} \textbf{A64}, 043405 (2001).

\bibitem{Thywissen}
J.H. Thywissen, M. Olsanii, G. Zabow, M. Drndi\'{c}, K.S. Johnson,
R.M. Westerwelt and M. Prentiss , \textit{Eur. Phys. J.}
\textbf{D7}, 361 (1999).

\bibitem{Reichel2}
J. Reichel, W. H\"ansel, P. Hommelhoff and T.W. H\"ansch:,
\textit{Appl. Phys.} \textbf{B72}, 81 (2001).

\bibitem{Yin}
J. Yin, W. Gao, J. Hu and Y. Wang, \textit{Opt. Commun.}
\textbf{206}, 99 (2002).

\bibitem{Hinds}
E. A. Hinds, C. J. Vale and M. G. Boshier, \textit{Phys. Rev.
Lett.} \textbf{86}, 1462 (2001).

\bibitem{Lau}
D.C. Lau, A.I. Sidorov, G.I. Opat, R.J. McLean, W.J. Rowlands and
P. Hannaford, \textit{Eur. Phys. J.} \textbf{D5}, 193 (1999).

\bibitem{Lee}
K. I. Lee, J. A. Kim, H. R. Noh and W. Jhe, \textit{Opt. Lett.}
\textbf{21}, 1177 (1996).

\bibitem{Schneble}
D. Schneble, H. Gauck, M. Hartl, T. Pfau and J. Mlynek, in
\textit{Proceedings of the International School of Physics
''Enrico Fermi'' Course CXL}, edited by M. Inguscio, S. Stringari,
and C.E. Wieman (IOS Press Amsterdam, 1999), pp. 469-490.

\bibitem{Zabov}
G. Zabow, M. Drndi\'{c}, J.H. Thywissen, K.S. Johnson, R.M.
Westervelt and M. Prentiss, \textit{Eur. Phys. J.} \textbf{D7},
351 (1999).

\bibitem{SAES}
SAES Getters S.p.A., 20151 Milano, Italy.

\bibitem{Fortagh}
J. Fortagh, A. Grossmann, T. W. H\"ansch and C. Zimmermann, {\it
J. Appl. Phys.} \textbf{84}, 6499 (1998).

\end{thebibliography}
\end{document}